\documentclass[11pt,a4paper]{article}

\usepackage[sort&compress,numbers]{natbib}
\usepackage{stackrel}
\usepackage{amssymb}
\usepackage{amsmath}
\input epsf 
\usepackage[pdftex]{graphicx}
\usepackage{graphics}
\usepackage{epsfig}
\usepackage{color}

\newcommand{\be}{\begin{equation}}
\newcommand{\ee}{\end{equation}}
\numberwithin{equation}{section}

\title{\bf{Tavis-Cummings models and their}\\
quasi-exactly solvable Schr\"odinger Hamiltonians}

\author{  
T. Mohamadian$^{1}$\footnote{tayebeh.mohamadian.k@gmail.com}, 
J. Negro$^2$\footnote{jnegro@fta.uva.es}, 
L.M. Nieto$^2$\footnote{luismiguel.nieto.calzada@uva.es}, and
H. Panahi$^1$\footnote{t-panahi@guilan.ac.ir},
\\ [2ex] 
\footnotesize \sl  $^1$Department of Physics, University of Guilan, Rasht 41635-1914, Iran.\\
\footnotesize \sl $^2$Departamento de F\'{\i}sica Te\'{o}rica, At\'{o}mica y \'{O}ptica, and IMUVA,\\
\footnotesize \sl Universidad de Valladolid, 47011 Valladolid, Spain.
}

\begin{document}

\maketitle

\begin{abstract}
We  study in detail the relationship between  the Tavis-Cummings Hamiltonian of quantum optics and a family of quasi-exactly solvable Schr\"odinger equations.
The connection   between them is stablished  through the biconfluent Heun equation. We  found that each invariant $n$-dimensional subspace of Tavis-Cummings Hamiltonian
corresponds either   to  $n$ potentials, each with one known solution, or   to one potential with $n$-known solutions. Among these Schr\"odinger potentials
 appear the quarkonium and the sextic oscillator.
\end{abstract}

\medskip


\medskip

%%%%%%%%%%%%%%%%%%%%%%%%%%%%%%%%%%%%%%%%%%%%%%%%%%%%%%%%%%%%%%%%%%%%%%%%%%%

\section{Introduction}

The main objective of this paper is to show  a direct method to connect
quantum optics Hamiltonians with quasi-exactly solvable (QES) one-dimensional Schr\"odinger equations
by considering the case of a trilinear Hamiltonian.
In principle, these two topics are quite far away since optical Hamiltonians make
use of operators corresponding to the radiation modes or to the interacting atoms
with a number of allowed transitions, while Schr\"odinger equations  describe  a (one dimensional, in this case) particle under an external potential.
However, the relation between the solutions of quantum optical Hamiltonians and Schr\"odinger wavefunctions has been   already  considered in some references (see \citep{bashir1995most,bogoliubov1996exact,lee2010polynomial}).
  Our purpose   in this work is to illustrate this connection in a clear and explicit way, including all the   relevant  details.

The quantum optical Hamiltonians we consider are modifications of the Jaynes-Cummings model \cite{jaynes1963comparison} in the rotating wave approximation (or RWA) describing the interaction of a one-mode quantum radiation field with a two-level atom. Here, we will  specially be concerned with the Dicke or Tavis-Cummings (TC) Hamiltonians 
which take into account the interaction of radiation with a population of two-level atoms
\cite{tavis1968exact}
(other cases such as three-level atoms can also be  considered, but it will not be done here).
With the help of the Schwinger representation we will write the TC Hamiltonian in  the somewhat  
more general form of a trilinear Hamiltonian \cite{walls1972nonlinear}, which has been
applied in many processes, like stimulated Raman and Brillouin effects or parametric
amplifiers   frequency conversion \citep{mishkin1969quantum,tucker1969quantum,walls1970quantum}. It is worth to mention that several methods were used to deal with this Hamiltonian: 
Bethe ansatz  \citep{bogolyubov2000algebraic,kundu2004quantum,lee2011exact}, 
algebraic approach  \cite{vadeiko2003algebraic,choreno2018algebraic}, 
quantum inverse method \cite{bogoliubov1996exact,rybin19981,bogoliubov2017time}, and 
strong radiation field \cite{kumar1980theory, kumar1981theory}.
Recently this trilinear Hamiltonian has  attracted much attention due to its  applications to study correlation functions \cite{abdalla2017linear, bogoliubov2017time} 
in    Circuit Quantum Electrodynamics Systems  \cite{fink2009dressed, romero2012ultrafast, xu2016dynamics, sun2017quantum,gong2018discrete}. In this paper we will show
how this trilinear quantum optical Hamiltonian is  associated with the biconfluent 
Heun equation (BHE), and  with various families of QES Schr\"odinger equations.

The organization of this work is as follows. In Section 2 we will introduce the TC
Hamiltonian in the form of a trilinear Hamiltonian, with its symmetries and invariant
subspaces labeled by  two integer parameters $(\ell,m)$. In Section 3 we will consider its differential realizations starting from the standard Fock-Bargman variables and propose a set of new variables in which the differential equation separates and becomes a BHE. Section 4 is devoted  to derive a transformation, characterized by label $b$, of the BHE into a Schr\"odinger equation.  The resulting potentials will be parametrized in terms of the initial modes, the $(\ell,m)$ subspaces and the transformation label $b$. Some examples are displayed in Section 5. Finally, some conclusions put an end to the paper.

%%%%%%%%%%%%%%%%%%%%%%%%%%%%%%%%%%%%%%%%%%%%%%%%%%%%%%%%%%%%%%%%%%%%%%%%%%%

\section{The Tavis-Cummings model: algebraic description}

The Tavis-Cummings or Dicke model describes the interaction of two-level atoms with a 
single mode of radiation field within the rotating wave approximation by means of
the following Hamiltonian, 
\begin{equation}\label{TCH}
H_{TC} = \omega_c c^+ c + \omega_c J_z + g(cJ_+ + c^+ J_-), 
\end{equation}
where $J_\pm,J_z$ are collective $su(2)$ operators for the atoms; $c,c^+$ are boson operators of the radiation field and $g$ is the coupling constant.
Now, we will use the Schwinger representation
\begin{equation}
J_+ = a^+b,\quad J_-= ab^+,\quad J_z = \frac12(a^+a - b^+b),
\end{equation}
 being $a,a^+$ and $b,b^+$ independent boson operators.
If besides this replacement we include some additional diagonal terms, then from (\ref{TCH}) we get the following trilinear Hamiltonian, in which three different modes of frequencies 
$\omega_1$, $\omega_2$, and $\omega_3$ interact with each other,
\be
H_{t}=\omega_a\, a^+ a+ \omega_b\, b^+ b+\omega_c\, c^+ c+ g(a^+ b c+a b^+ c^+) \, .
\ee 
The term with $g\neq 0$ now represents the interaction between these three modes. It is convenient to use a simplified version of this Hamiltonian, obtained by dividing the previous one by the coupling constant $g$:
\be\label{tcham}
H=\frac{H_{t}}{g}=\omega_1\, a^+ a+ \omega_2\, b^+ b+\omega_3\, c^+ c+ a^+ b c+a b^+ c^+,
\ee 
with $\omega_1={\omega_a}/g$, $\omega_2={\omega_b}/g$, $\omega_3={\omega_c}/g$.
This system has a set of three independent and commuting symmetry operators,
including the Hamiltonian $H$,
\begin{equation}\label{symmetries}
\{ \, H,\  L=N_a+N_b,\  M=N_a+N_c \, \},
\end{equation}
where $N_a= a^+ a$, $N_b= b^+ b$  and  $N_c=  c^+ c$
are the number operators of the corresponding modes. As the system
has three degrees of freedom, we can say that it is  completely
integrable \cite{alvarez2002quasi, miller2013classical}.

Let us consider now the Fock space representation of the system in terms of the number of photons of each frequency, 
$\{ |n_a,n_b,n_c\rangle;\, n_a,n_b,n_c\in \mathbb{N} \}$. 
The symmetries $L$ and $M$ allow us to define subspaces
${\cal W}_{\ell,m}$, characterized by their respective simultaneous eigenvalues $\ell, m$, which have the following properties:
\begin{itemize}
\item
 The subspaces ${\cal W}_{\ell,m}$ are finite dimensional and their corresponding dimensions are given by 
 $d={\rm min}(\ell,m)+1$.  Each subspace (for instance if $\ell\geq m$) is generated by the following basis vectors $|n_a,n_b,n_c\rangle$:
\be\label{subspace}
\begin{tabular}{|c|c|c|}
\hline
               &  $\ell\geq m$ &    \\
\hline
\hline
         $n_a$       &  $n_b$  &  $n_c$   \\
\hline
$0$ &     $\ell$      &      $m$   \\
\hline
 $1$   &     $\ell-1$      &      $m-1$   \\
\hline
 $2$  &     $\ell-2$      &      $m-2$   \\
\hline
$\dots$  &     $\dots$      &      $\dots$   \\
\hline
$m$  &     $\ell-m$      &      $0$   \\
\hline
\end{tabular}
%\qquad\qquad\qquad\qquad
%\begin{tabular}{|c|c|c|}
%\hline
%               &  $\ell\leq m$ &    \\
%\hline
%\hline
%         $n_a$       &  $n_b$  &  $n_c$   \\
%\hline
%$0$ &     $\ell$      &      $m$   \\
%\hline
% $1$   &     $\ell-1$      &      $m-1$   \\
%\hline
% $2$  &     $\ell-2$      &      $m-2$   \\
%\hline
%$\dots$  &     $\dots$      &      $\dots$   \\
%\hline
%$\ell$  &     $0$      &      $m-\ell$   \\
%\hline
%\end{tabular}
\ee

\item
The subspaces ${\cal W}_{\ell,m}$ are invariant under the Hamiltonian $H$. 
Therefore, the Hamiltonian can be restricted to each
invariant subspace ${\cal W}_{\ell,m}$, this restriction  will be denoted by $H_{\ell,m}$. Then,  we can look for the spectrum of each $H_{\ell,m}$, in this way
the eigenvalue problem for $H$ has been reduced to a sequence of finite dimensional problems. This is the origin of  the  quasi-exact solvability associated to the Tavis-Cummings (or to the
trilinear) Hamiltonian.
\end{itemize}

\subsection{Examples \label{examples}}

Among the ${\cal W}_{\ell,m}$ there is an infinity of $d$-dimensinal subspaces. For example those of the type ${\cal W}_{\ell,0}$ and ${\cal W}_{0,m}$ are one-dimensional, those of the type ${\cal W}_{\ell\geq 1,1}$ and ${\cal W}_{1,m\geq1}$ are two-dimensional, etc.
We consider now two simple examples that later on  will be worked out in the differential realisations.

\begin{itemize}

\item{ $\ell=m = 1$}

\noindent
The two dimensional subspace ${\cal W}_{1,1}$ in this case is generated by the vectors
\begin{equation}\label{W11}
\{ |1,0,0 \rangle,\, |0,1,1 \rangle \}.
\end{equation}
The matrix representation of the Hamiltonian in this subspace will be given by
\begin{equation}\label{h11}
H_{1,1}=\left(
\begin{array}{ccc}
\omega_1&1\\
1&\omega_1+\omega_2
\end{array}
\right) \ .
\end{equation}

\item{$\ell=3,$ $m = 2$}

\noindent
Here, the subspace ${\cal W}_{3,2}$ is generated by the vectors
\begin{equation}\label{W32}
\{ |2,1,0 \rangle,\, |0,3,2 \rangle,\, |1,2,1\rangle\},
\end{equation}
and the Hamiltonian in this subspace (taking the above basis) is:
\begin{equation}
\label{h32}
H_{3,2}=\left(
\begin{array}{ccc}
2\omega_1+\omega_2&0&2\\
0&3\omega_2+2\omega_3&\sqrt{6}\\
2&\sqrt{6}& \omega_1+2\omega_2+\omega_3
\end{array}
\right) \ .
\end{equation}
\end{itemize}
Remark that the restricted Hamiltonians $H_{\ell,m}$ are explicitly
Hermitian and therefore  they can be  diagonalized with real eigenvalues.

%%%%%%%%%%%%%%%%%%%%%%%%%%%%%%%%%%%%%%%%%%%%%%%%%%%%%%%%%%%%%%%%%%%%%%%%%%%

\section{Differential realizations  of the Tavis-Cummings model}

The simplest differential realization  of the trilinear form of the Tavis-Cummings Hamiltonian (\ref{tcham}), is the Fock-Bargmann representation in terms of three complex variables:
\be
a=\partial_{z_1}, \quad a^+=z_1, \qquad 
b=\partial_{z_2}, \quad b^+=z_2, \qquad 
c=\partial_{z_3}, \quad c^+=z_3.
\ee
Using these variables, the Hamiltonian \eqref{tcham} adopts the differential form
\be
H_{T}=\omega_1 z_1\, \partial_{z_1} + \omega_2 z_2\, \partial_{z_2} +\omega_3 z_3\, \partial_{z_3} + 
z_1\, \partial_{z_2} \partial_{z_3} +z_2\, z_3\,  \partial_{z_1},
\ee
while the symmetry operators $L$ and $M$   are first order differential operators
\be
L= z_1\, \partial_{z_1} +  z_2\, \partial_{z_2}, \qquad 
M= z_1\, \partial_{z_1} +  z_3\, \partial_{z_3}.
\label{nmzeta}
\ee
In this realization the Fock states are given by the monomials 
\begin{equation}
|n_a,n_b,n_c\rangle= \frac{(z_1)^{n_a}(z_2)^{n_b}(z_3)^{n_c}}{\left(n_a!\,n_b!\,n_c!\right)^{1/2}}\, .
\end{equation}

Taking into account the symmetries \eqref{symmetries} of the system, we are going to make a transformation from the set of variables $(z_1,z_2,z_3)$ to another set $(r,s,t)$ so that the Hamiltonian can be simplified. These new variables are  characterized by the 
condition that the operators $L$ and $M$ must
depend only on one variable; here we will choose $L$ to be a function of $s$ and $M$ a function
of $t$.
Then, after straightforward computations, we find the following relationship between the two sets of variables
\be\label{changevar}
r=\frac{c z_1}{z_2z_3},\quad
s=\frac{z_1}{z_3}\, f(z_2),\quad
t=\frac{z_1}{z_2}\, g(z_3),
\ee
where $c$ is a constant and $f$, $g$ are arbitrary functions.
Without loss of generality, we can take these functions  as $f(z_2)=g(z_3)=1$. Hence, we can rewrite the variables $(z_1,z_2,z_3)$ in terms of $(r,s,t)$ as follows:
\be
z_1= {c} \, \frac{s t}{r}, \quad
z_2= {c}  \, \frac{s}{r} ,\quad
z_3= {c}  \,\frac{t}{r}.
\ee
 Therefore, partial derivatives transform as
\begin{eqnarray}
&&\frac{\partial}{\partial z_1}= \frac{1}{c}\left( 
\frac{r}{t}\frac{\partial}{\partial s}+  \frac{r}{s} \frac{\partial}{\partial t}+ \frac{r^2}{st}\frac{\partial}{\partial r}  \right), 
\\
&&
\frac{\partial}{\partial z_2}=  -\frac{1}{c}\left( \frac{t r}{s} \frac{\partial}{\partial t} +\frac{r^2}{s} \frac{\partial}{\partial r}   \right), 
\\
&&
\frac{\partial}{\partial z_3}= -\frac{1}{c}\left(  \frac{rs}{t} \frac{\partial}{\partial s} +  \frac{r^2}{t} \frac{\partial}{\partial r}  \right).
\end{eqnarray}
The three commmuting operators $H$, $L$ and $M$ have the following expressions in terms of the new variables $(r,s,t)$:
\begin{eqnarray}
H \!\!&\!=\!&\!\! \frac{r^3}c\, \frac{\partial^2}{\partial r^2}+\frac{rst}c\, \frac{\partial^2}{\partial s \partial  t}+ \frac{tr^2}c\, \frac{\partial^2}{\partial r \partial  t}+\frac{r^2 s}c \, \frac{\partial^2}{\partial r \partial  s} 
\\[1.5ex]
&& + \left( \frac{r^2}c +(\omega_1-\omega_2-\omega_3)r+ c \right)  \frac{\partial}{\partial r} + ((\omega_1-\omega_3)r+c) \frac{s}{r} \frac{\partial}{\partial s} + ((\omega_1-\omega_2)r+c) \frac{t}{r} \frac{\partial}{\partial t} .   
\nonumber
\\[2.ex]
M\!\!&\!=\!&\!\! t\, \frac{\partial}{\partial t}, \qquad L=s\, \frac{\partial}{\partial s} .
\end{eqnarray}

Now, we  can look for solutions  $\Psi(r,s,t)$ which are simultaneous eigenfunctions of
the symmetries $L$ and $M$, i.e., they belong to the subspace ${\cal W}_{\ell,m}$ characterized by
\be
M \Psi(r,s,t)=m \Psi(r,s,t), \qquad L\Psi(r,s,t) = \ell \Psi(r,s,t).
\ee\label{sepfun}
Then, the wavefunction factorizes as 
\begin{equation}\label{separated}
\Psi(r,s,t)=s^\ell\, t^m\, \psi(r)\, ,
\end{equation}
 and the equation for the Hamiltonian $H$ with eigenvalue $E$  in the subspace ${\cal W}_{\ell,m}$ admits separation in these variables,  leading to the following linear second-order ordinary differential equation in terms of the variable $r$:
\begin{eqnarray}\label{firsheun}
&& \frac{d^2\psi}{dr^2}+\left( \frac{c^2}{r^3}+\frac{c(\omega_1-\omega_2-\omega_3)}{r^2}+\frac{m+\ell+1}{r} \right) \frac{d\psi}{dr}+ \nonumber
\\
&& \qquad \, + \left( \frac{(m+\ell)c^2}{r^4}+\frac{c(m(\omega_1-\omega_2)+\ell(\omega_1-\omega_3)-E)}{r^3}+\frac{m \ell}{r^2} \right) \psi=0.
\end{eqnarray}

We would like the solutions  of (\ref{separated}) to be polynomials in all the new variables, so that  $\psi(r) =  a_0 +a_1 r+a_2 r^2+\cdots a_N r^N$, then if we replace in  (\ref{separated}) the old variables, we have
\begin{equation}
\Psi(z_1,z_2,z_3)=\frac{z_1^{m+\ell}}{z_2^m z_3^\ell}\, 
\sum_{k=0}^N a_k \left(\frac{c z_1}{z_2 z_3} \right)^k . 
\end{equation}
However, this wavefunction should also be a polynomial in the old variables, but in its present form the denominators
exclude this possibility. Therefore, in order to solve this  difficulty we must change from  the variable $r$ to its inverse $\rho = 1/r$, so that the above relation  becomes
\begin{equation}
\Psi(z_1,z_2,z_3)=\frac{z_1^{m+\ell}}{z_2^m z_3^\ell}\, 
 \sum_{k={\rm max}(\ell,m)}^{\ell+m} a_k \left(\frac{z_2 z_3}{c z_1}\right)^k,
\end{equation}
or  in terms of the variables $\rho, s, t$:
\begin{equation}\label{stack}
\Psi(\rho,s,t)= s^\ell t^m \rho^{{\rm max}(\ell,m)}\, 
\sum_{n=0}^{N'}
%\substack{\ell+m- \\
%{\rm max}(\ell,m)}} 
b_n  \rho^n \, ,\quad N'=\ell+m- {\rm max}(\ell,m),
\end{equation}
having thus an acceptable polynomial form.
Hence, in terms of $\rho=1/r$, equation \eqref{firsheun} transforms into
\begin{eqnarray}\label{secondheun}
&& \frac{d^2\psi}{d\rho^2}-\left( \frac{m+\ell-1}{\rho}+ c(\omega_1-\omega_2-\omega_3) +c^2 \rho \right) \frac{d\psi}{d\rho}+ \nonumber
\\[1.5ex]
&& \qquad \, + \left(\frac{m \ell}{\rho^2} + \frac{c(m(\omega_1-\omega_2)+\ell(\omega_1-\omega_3)-E)}{\rho}+ (m+\ell)c^2 \right) \psi=0\,.
\end{eqnarray}
This equation has a regular singularity at the origin and an irregular singularity
of rank two at infinity, which coincides with the singularities of the biconfluent Heun equation (BHE). Indeed, if we transform \eqref{secondheun} using the change of function $\psi(\rho)= \rho^k \varphi(\rho)$, where $k= {\rm max}(\ell,m)$, as suggested by (\ref{stack}), we get
\begin{eqnarray}\label{BCE}
&\hskip-0.5cm &
\frac{d^2 \varphi}{d\rho^2} 
 +
\left( \frac{1+2k-m-\ell}{\rho}- c(\omega_1-\omega_2-\omega_3) -c^2 \rho \right) 
\frac{d \varphi}{d\rho}+ \left(\frac{(m-k)(\ell-k)}{\rho^2} + \right. 
\\ [1ex]
&\hskip-0.5cm & \qquad  \left.+ \frac{c(m(\omega_1-\omega_2)+\ell(\omega_1-\omega_3)-E-k(\omega_1-\omega_2-\omega_3))}{\rho}+ (m+\ell-k)c^2 \right) \varphi=0. 
\nonumber
\end{eqnarray}
From \eqref{BCE}, we can see that if $k=\ell$ or $k=m$ this equation has the form of a 
BHE \cite{Heun}, with general form
\begin{equation}\label{BCEdef}
y''+\left( \frac{1+\alpha}x +\beta-2x \right)y'+\left(-\frac{\delta+(1+\alpha)\beta}{2x}+\gamma-\alpha-2   \right)y=0.
\end{equation}
Let us observe that comparing \eqref{BCE} and \eqref{BCEdef} we must have  $c=\pm \sqrt{2}$.
Indeed, the coefficientes $\alpha, \beta,\gamma$ should be of the form:
\begin{enumerate}
\item
If $m\geq \ell$, then $k=m$:
\begin{equation}
\begin{array}{l}
\alpha=m-\ell, \qquad \beta= -c(\omega_1-\omega_2-\omega_3),\qquad \gamma= m+\ell+2, 
\\[1.5ex] 
\delta= c \left( m (\omega_1-\omega_2-3\omega_3)- \ell (3\omega_1-\omega_2-3\omega_3)+ (\omega_1-\omega_2-\omega_3+2E)\right) .
\end{array}
\end{equation}
\item
If $\ell\geq m$, then $k=\ell$. In this case the identification of the parameters
$\alpha, \beta, \gamma$ and $\delta$ is the same as above, but interchanging $\omega_2$
with $\omega_3$, and $\ell$ with $m$.
\end{enumerate}

%%%%%%%%%%%%%%%%%%%%%%%%%%%%%%%%%%%%%%

\section{From the biconfluent Heun equation to a Schr\"odinger equation}

In this Section we will transform the BHE \eqref{BCE} into a more familiar Schr\"odinger-type equation, that will be analyzed in detail.

If we  consider the value $c=\sqrt{2}$ (the other sign choice will be discussed later) in \eqref{BCE}, and make the change of dependent and independent variables \cite{zaslavskii1990effective,gomez2005quasi}
\begin{equation}\label{wf}
\varphi(\rho)= e^{-W(x)}\, \chi (x),\qquad x = x(\rho),
\end{equation}
we have the possibility to choose the functions $W(x)$ and $x(\rho)$ in such a way 
that $\chi(x)$ satisfy a Schr\"odinger equation, with eigenvalue   $\lambda=0$:
\begin{equation}\label{sch}
 -\chi''+V(x)\chi =\lambda \chi=0 .
\end{equation} Indeed, if $W(x)$ satisfies the equation
\begin{equation}
\ddot{x}  -2 \left( \dot{x}\right)^2 \frac{d W}{d x}+
\frac{d x}{d\rho}\left( \frac{2k-m-n+1}{\rho} -\sqrt{2}(\omega_1-\omega_2-\omega_3) -2\rho \right)=0,
\end{equation}
where $\dot{x}=dx/d\rho$, the potential $V(x)$ introduced in \eqref{sch} has the form
\begin{eqnarray}
V(x) \!\!&  \!\!= \!\!&  \!\! \frac{\ddot{x}}{\dot{x}^2} \frac{d W}{d x}-\left( \frac{d W}{d x}\right)^2+ 
\frac{d^2 W}{d x^2}+\frac1{\dot{x}}\, \frac{d W}{d x} \left[ \frac{2k-m-\ell+1}{\rho} -\sqrt{2} 
(\omega_1{-}\omega_2{-}\omega_3) {-} 2\rho \right]  \nonumber
\\[1.5ex]
\!\!&  \!\! \!\!&   \!\! \!\!
 - \frac1{\dot{x}^2}\left[ \frac{\sqrt{2} \left( \ell (\omega_1-\omega_3)+  m (\omega_1-\omega_2)-E-k (\omega_1-\omega_2-\omega_3) \right)}{\rho}  +2(m+\ell-k)  \right].
\end{eqnarray}
Henceforth, we choose the dependence between $x$ and $\rho$ in the form $x= \rho^b$, $b>0$,  and therefore the potential (denoted from now on by $V_b(x)$) and the wavefunction $\chi(x)$ will have the following expressions:
\begin{eqnarray}
&&V_b(x) =  \left( -\frac14+\frac{(1+B)^2-4k-4kB+4k^2}{4b^2}  \right) x^{-2}-\left( \frac{AB+2D}{2b^2} \right) x^{-2+\frac1b}  
\nonumber
\\ [1ex]
&& \qquad \qquad - \left( \frac{A^2+4B+4G-4}{4b^2} \right) x^{-2+\frac2b}
 +  \frac{A}{b^2}\ x^{-2+\frac3b}
 +  \frac{1}{b^2}\ x^{-2+\frac4b},
 \label{pot}
\\ [2ex]
\label{chi}
&&\chi(x)= 
x^{-(1-b+B-2k)/(2b)}\exp\left[-\frac1{2} x^{1/b}\left(A+ x^{1/b}\right)\right]\, \varphi(\rho(x)) ,
\end{eqnarray}
where
\begin{equation}
\begin{array}{l}
A= \sqrt{2} (\omega_1-\omega_2-\omega_3), \quad B=m+\ell-1, \quad G=2(m+\ell), 
\\[1.ex]
D=\sqrt{2} (m (\omega_1-\omega_2) + \ell (\omega_1-\omega_3)-E).
\end{array}
\end{equation}
Notice that for all the values of the parameters $\ell,m \in \mathbb N$ and $b>0$,
the power of $x$ in front of (\ref{chi}) is positive, while the dominant term in the
exponential is negative. This means that in all cases the resulting solutions
$\chi(x)$ will be square integrable wavefunctions such that 
\begin{equation}
\lim_{x\to 0} \chi(x) =0, \qquad \lim_{x\to\infty} \chi(x) =0,
\end{equation}
which make them physically acceptable.

Now we are interested in removing some of the terms in the potential (\ref{pot}),  in order to have a simpler form. To meet this goal we can select some specific values of $b$:
\begin{itemize}
\item
If $b=1$, the potential is 
\begin{eqnarray}
V_{1}(x)\!\!&\!=\!&\!\!\frac{\sqrt{2}(2E+(1-3m-3n)\omega_1+(-1+3m+n)\omega_2+(-1+m+3n)\omega_3)}{2x} \nonumber \\
\!\!&\! \!&\!\!
+ \frac{(m-\ell-1)(m-\ell+1)}{4x^2}
%\nonumber\\  \!\!&\! \!&\!\!
+2-3(m+n) -\frac{(-\omega_1+\omega_2+\omega_3)^2}{2} \nonumber \\
\!\!&\! \!&\!\!
 +\sqrt{2}(\omega_1-\omega_2-\omega_3)\,x+x^2,  \label{potb1}
\end{eqnarray}
and the wavefunction 
\begin{equation}
\chi(x)= x^{-(B-2k)/2}\exp\left[-\frac1{2}  {x}  \left(A+  {x} \right)\right]\, \varphi(\rho(x)) \,.
\end{equation}
Remark that this potential corresponds to a combination of Coulomb, oscillator and linear   potentials, together with a centrifugal term. This potential is used to describe the quarkonium \cite{karayer2018solution,Ovsiyuk} and a two-electron quantum dot 
\cite{caruso2014solving}.

\item
If $b=1/2$, the potential becomes 
\begin{equation}
\begin{array}{l}
V_{1/2}(x) =  \displaystyle \frac{(2m-2\ell-1)(2m-2\ell+1)}{4 x^2} +4\sqrt{2}  (\omega_1-\omega_2-\omega_3)\, x^4 +4x^6  
\\[1.5ex]
\qquad\quad
-(-8 + 12(m+\ell) +2(\omega_1-\omega_2-\omega_3)^2)\, x^2   \\[1.5ex]
\qquad\quad
+2\sqrt{2}[-(3m+3\ell-1)\omega_1+(\ell+3m-1)\omega_2+(3\ell+m-1)\omega_3] 
-\varepsilon(E)
\\[1.5ex]
\qquad\quad
:=  \tilde{V}_{1/2}(x) -\varepsilon(E),  
\end{array}
  \label{potb12}
\end{equation}
where the function $\varepsilon(E)$ does not depend on $x$
\begin{equation}
\varepsilon(E) = -4\sqrt{2}\,E
\end{equation} 
and the form of the wavefunction is
\begin{equation}\label{eigenchi}
\chi(x)= x^{-(B-2k+\frac{1}{2})} \exp\left[-\frac1{2}  {x^2}  \left(A+  {x^2} \right)\right]\, \varphi(\rho(x)) \,.
\end{equation}
In this case, the potential consists of a sextic potential plus the centrifugal term.
This potential has been studied in several references, where it was derived by other methods \cite{lee2010polynomial, sobhani2017analytical, quesne2018quasi}. It is remarkable that the term $\varepsilon(E)$ in the potential $V_{1/2}(x)$ given by (\ref{potb12}) can be moved to the right hand side of its corresponding Schr\"odinger equation (\ref{sch}), so that  we are faced with an equivalent Schr\"odinger equation for the displaced potential $\tilde{V}_{1/2}(x)$ with eigenvalue $\varepsilon(E)$ and eigenfunction $\chi(x)$ given in \eqref{eigenchi}.

\item
If $b=3/2$ the potential is 
\begin{eqnarray}
V_{3/2}(x)\!\!&\!=\!&\!\! \frac{(2m-2\ell-3)(2m-2\ell+3)}{36\, x^2} +\frac{4}{9}\, x^{2/3} +
\frac{4\sqrt{2}}{9}\, (\omega_1-\omega_2-\omega_3) \nonumber \\
\!\!&\! \!&\!\!
+\frac{2\sqrt{2}(2E-(3m+3\ell-1)\omega_1+(\ell+3m-1)\omega_2+(3\ell+m-1)\omega_3)}{9 \, x^{4/3}} \nonumber\\
\!\!&\! \!&\!\!
-\frac{(-8 + 12(m+\ell) +2(\omega_1-\omega_2-\omega_3)^2)}{9\, x^{2/3}}, 
\label{potb32}
\end{eqnarray}
and the corresponding wavefunction is
\begin{equation}
\chi(x)= x^{-(-\frac12+B-2k)/3}\exp\left[-\frac1{2}    {x}^{2/3}\left(A+ {x}^{2/3}\right) \right]\, \varphi(\rho(x)) \,.
\end{equation}

\item
If $b=2$ the potential is 
\begin{eqnarray}
V_2(x)\!\!&\!=\!&\!\! \frac{(m-\ell-2)(m-\ell+2)}{16 x^2} +\frac1{4} +
\frac{\sqrt{2}(\omega_1-\omega_2-\omega_3)}{ 4 \sqrt{x} } \nonumber\\
\!\!&\! \!&\!\!  
+\frac{2\sqrt{2}(2E-(3m+3\ell-1)\omega_1+(\ell+3m-1)\omega_2+(3\ell+m-1)\omega_3)}{16 \, x^{3/2}} \nonumber\\
\!\!&\! \!&\!\!
-\frac{(-8 + 12(m+\ell) +2(\omega_1-\omega_2-\omega_3)^2)}{x},  \label{potb2}
\end{eqnarray}
while the wavefunction is
\begin{equation}
\chi(x)= x^{-(-1+B-2k)/4}\exp\left[-\frac1{2}    {x}^{1/2}\left(A+ {x}^{1/2}\right)\right]\, \varphi(\rho(x)) \,.
\end{equation}
This potential $V_2(x)$ includes a long range term
$V(x) = -\alpha/\sqrt{x}$ that has been studied in \citep{karayer2018solution}.
Choosing appropriate values of the parameters, the term $x^\frac{3}{2}$ can be eliminated.

\end{itemize}

Notice that the potentials $V_1(x), V_{3/2}(x)$ and $V_2(x)$ include the eigenvalues $E$  of our original trilinear Hamiltonian as an intrinsic part of the parameters, so that the energy of the Schr\"odinger equation (\ref{sch}) must be zero, $\lambda=0$. In other words, for these values of $b$ we get a list of potentials with only one known wavefunction corresponding to the energy $\lambda=0$.

For the negative value $c= -\sqrt{2}$, we obtain the same equation (\ref{secondheun}) as for $c= +\sqrt{2}$, but replacing $\rho$ by $-\rho$.
This leads us to another set of potentials where we must take into account
that the variable $x$ is related to $\rho$ in the form
$x = (-\rho)^b$ (instead of $x = \rho^b$).

%%%%%%%%%%%%%%%%%%%%%%%%%%%%%%%%%%%%%%%%%%%%%%%%%%%%%%%%%%%%%%%%%%%%%%%%%%%

\section{Two examples: the quarkonium and the sextic potential}

From now on, in order to illustrate the general results previously obtained, we will concentrate on two cases mentioned in Section~\ref{examples},
(i) $m=\ell=1$, and (ii) $m=2$, $\ell=3$.
We will also restrict  to the cases $V_1(x)$ and $V_{1/2}(x)$, because they represent the  potentials of well known physical systems: the quarkonium potential, the two-electron quantum dot and the sextic oscillator.
 
%If $b=1$, the potential \eqref{potb1} is such that the parameter $E$ (the original energy of our system) appears inside the potential, $V_1(x,E)$. In this case, the corresponding Schr\"odinger equation can be interpreted as providing for the eigenvectors corresponding to zero eigenvalue of the potentials $V_1(x,E_\ell)$, where $E_\ell$ are the eigenvalues of the original problem under study.
%
%If $b=1/2$,  the potential \eqref{potb12} is such that the parameter $E$ can be separated from it considering a rescaled ptential $V^*_{1/2}(x)=V_{1/2}(x,E) -4\sqrt{2}E$, whose eigenvalues are $\lambda=-4\sqrt{2}E$.

\subsection{Case $m=\ell=1$}

The two dimensional subspace in this case is ${\cal W}_{1,1}$, given by \eqref{W11} and the matrix representation of the Hamiltonian $H_{1,1}$ is \eqref{h11}. If for simplicity we choose $\omega_1=\omega_2=\omega_3=1$, the eigenvalues and eigenfunctions corresponding to this two by two matrix $H_{1,1}$ are
\be
|\psi_p \rangle= \gamma_{p,1} |1,0,0 \rangle+ \gamma_{p,2} |0,1,1 \rangle ,
\quad p=1,2,
\ee
with
\be\label{e12}
\begin{tabular}{|c|c|r|r|}
\hline 
&&& \\ [-1.9ex]
  $p$   &   $E_p$        & $\gamma_{p,1}$\phantom{\quad} & $\gamma_{p,2}$\phantom{\quad}  \\ [0.9ex]
\hline 
&&& \\ [-1.9ex]
1& $\frac{3+\sqrt{5}}2$ \quad &     $\sqrt{\frac{5-\sqrt{5}}{10}}$ \quad     &      $\sqrt{\frac{5+\sqrt{5}}{10}}$  \quad \\ [0.9ex]
\hline 
&&& \\ [-1.9ex]
2& $\frac{3-\sqrt{5}}2$ \quad &     $-\sqrt{\frac{5+\sqrt{5}}{10}}$   \quad   &      $\sqrt{\frac{5-\sqrt{5}}{10}}$ \quad  \\[0.9ex]
\hline
\end{tabular}
\ee
We can write those eigenfunctions in terms of the variables $\rho, s, t$ 
(see (\ref{stack})) as follows:
\be
\Psi_p(\rho,s,t) =  s   t   \rho \left( \sqrt{2}\, \gamma_{p,1} + 2 \gamma_{p,2}\ \rho \right) =  s t \rho \, \varphi_p(\rho).
\ee
The wavefunction is
\be
\chi_p(x)= e^{W(x)} \left( \sqrt{2}\, \gamma_{p,1} + 2 \gamma_{p,2}\ \rho(x) \right) .
%=  x^{-(1-b+B-2k)/(2b)}\exp\left[-\frac1{2}  \left( {x}  \right)^{1/b}\left(A+ \left( {x} %\right)^{1/b}\right)\right]\left( \sqrt{2}\, \gamma_{p,1} + 2 \gamma_{p,2}\ \rho \right).
\ee

\begin{figure}[ht]
\centering
\includegraphics[width=0.47\textwidth]{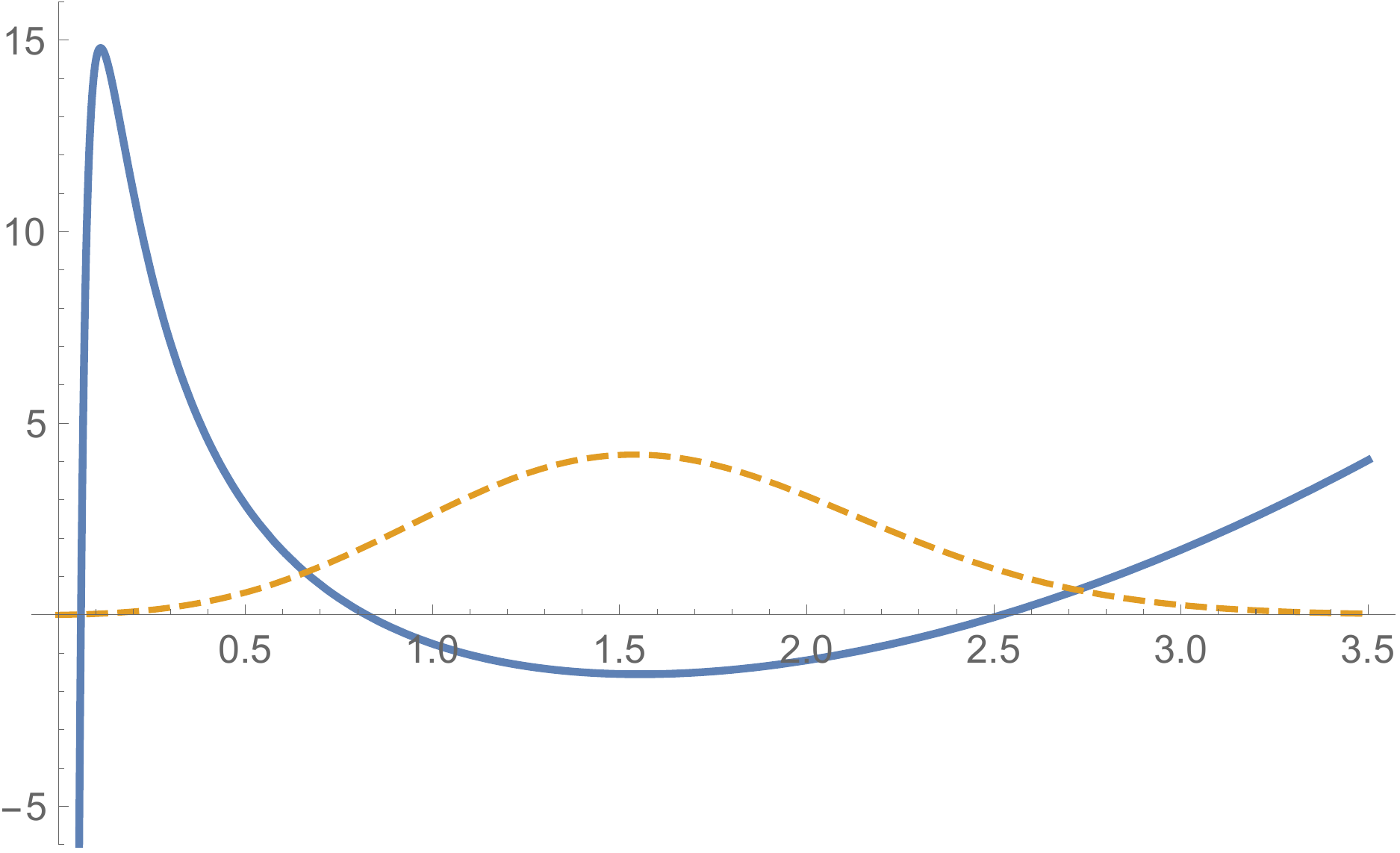}
\qquad
\centering
\includegraphics[width=0.47\textwidth]{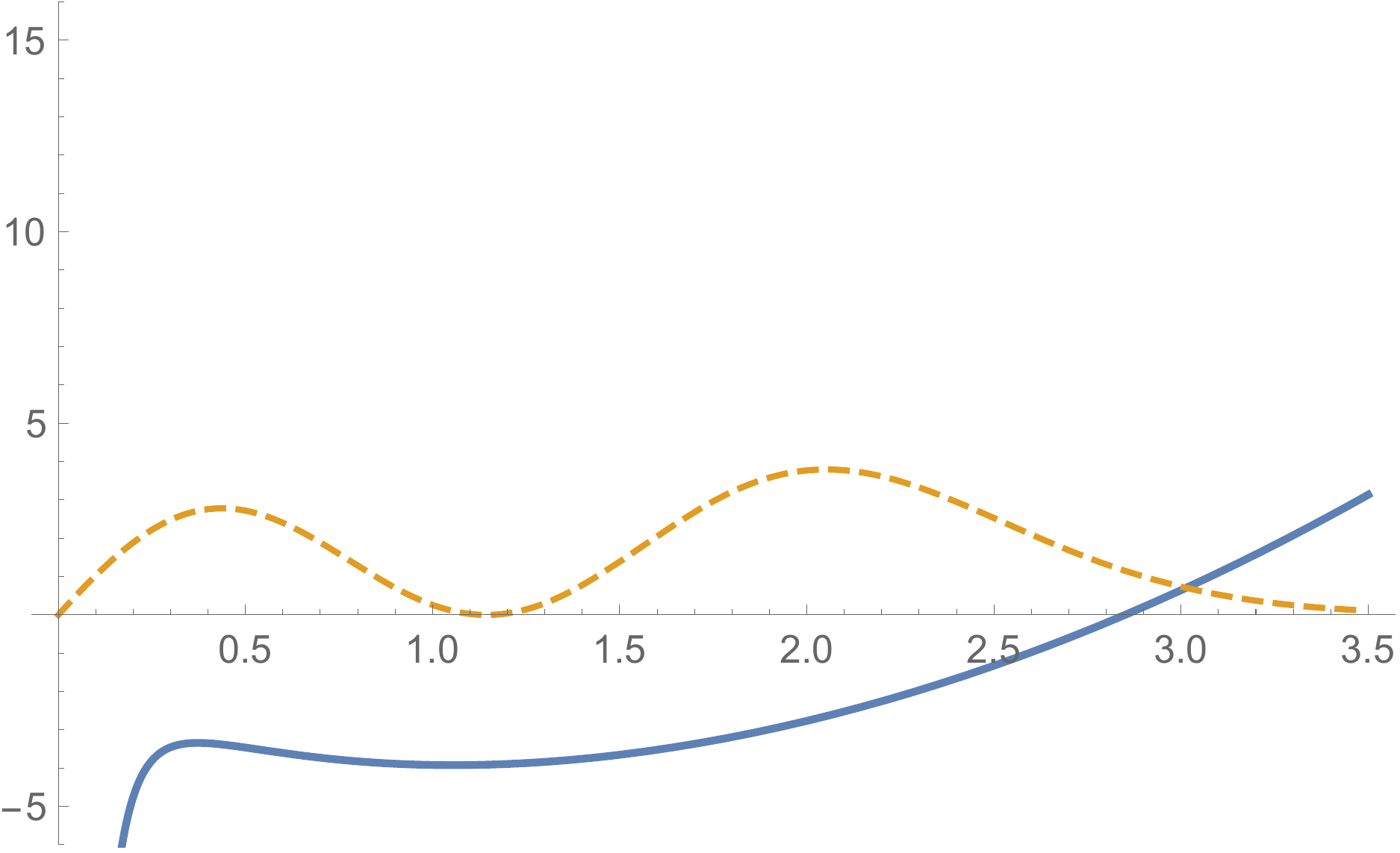}
\caption{The potential $V_1(x)$ (solid line) and probability density function $|\chi_p(x)|^2$ (dashed line) for $E_1=(3+\sqrt{5})/2$ (left) and $E_1=(3-\sqrt{5})/2$ (right).}
\label{fig12}
\end{figure}

\begin{figure}[ht]
\centering
\includegraphics[width=0.47\textwidth]{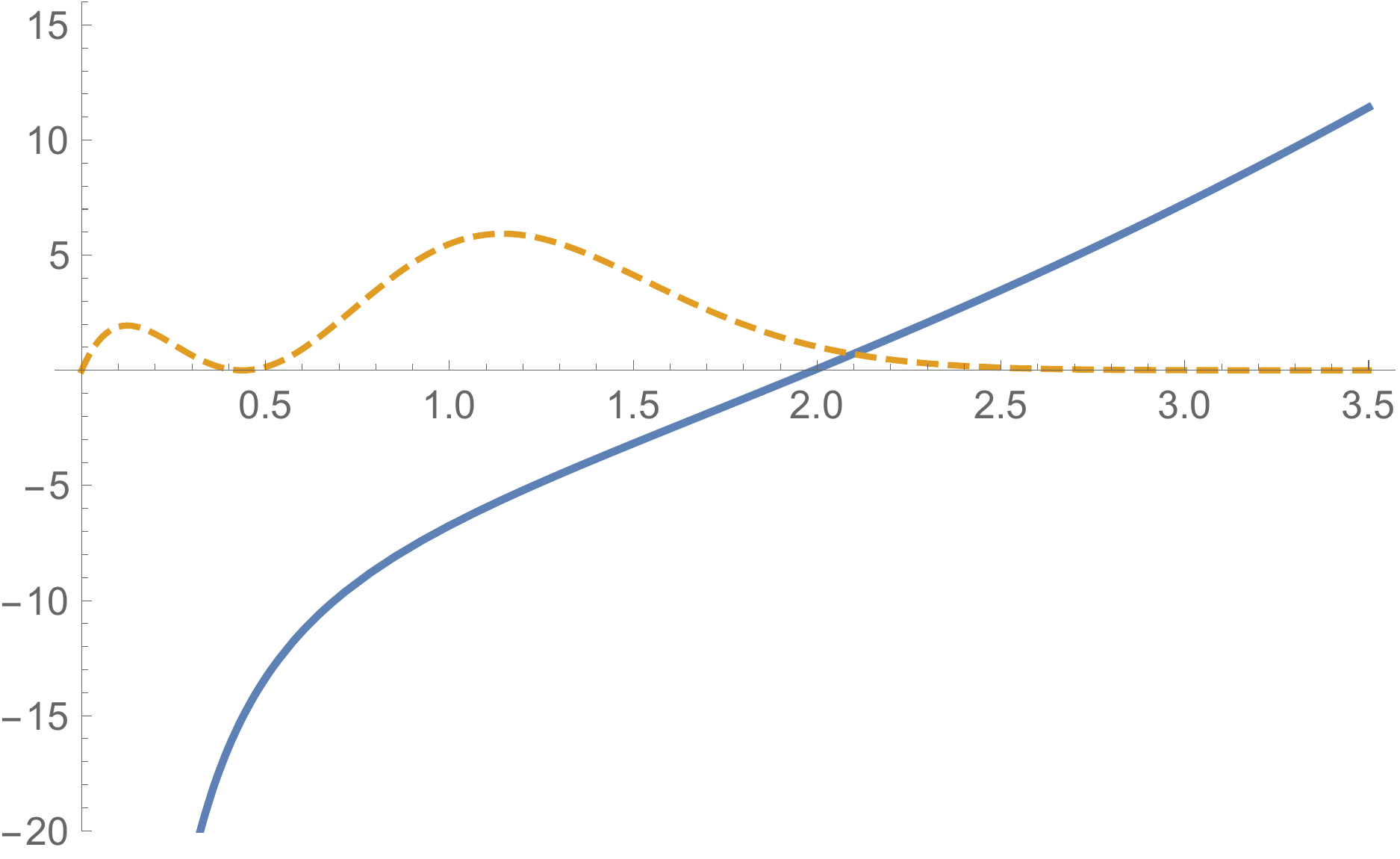}
\qquad
\centering
\includegraphics[width=0.47\textwidth]{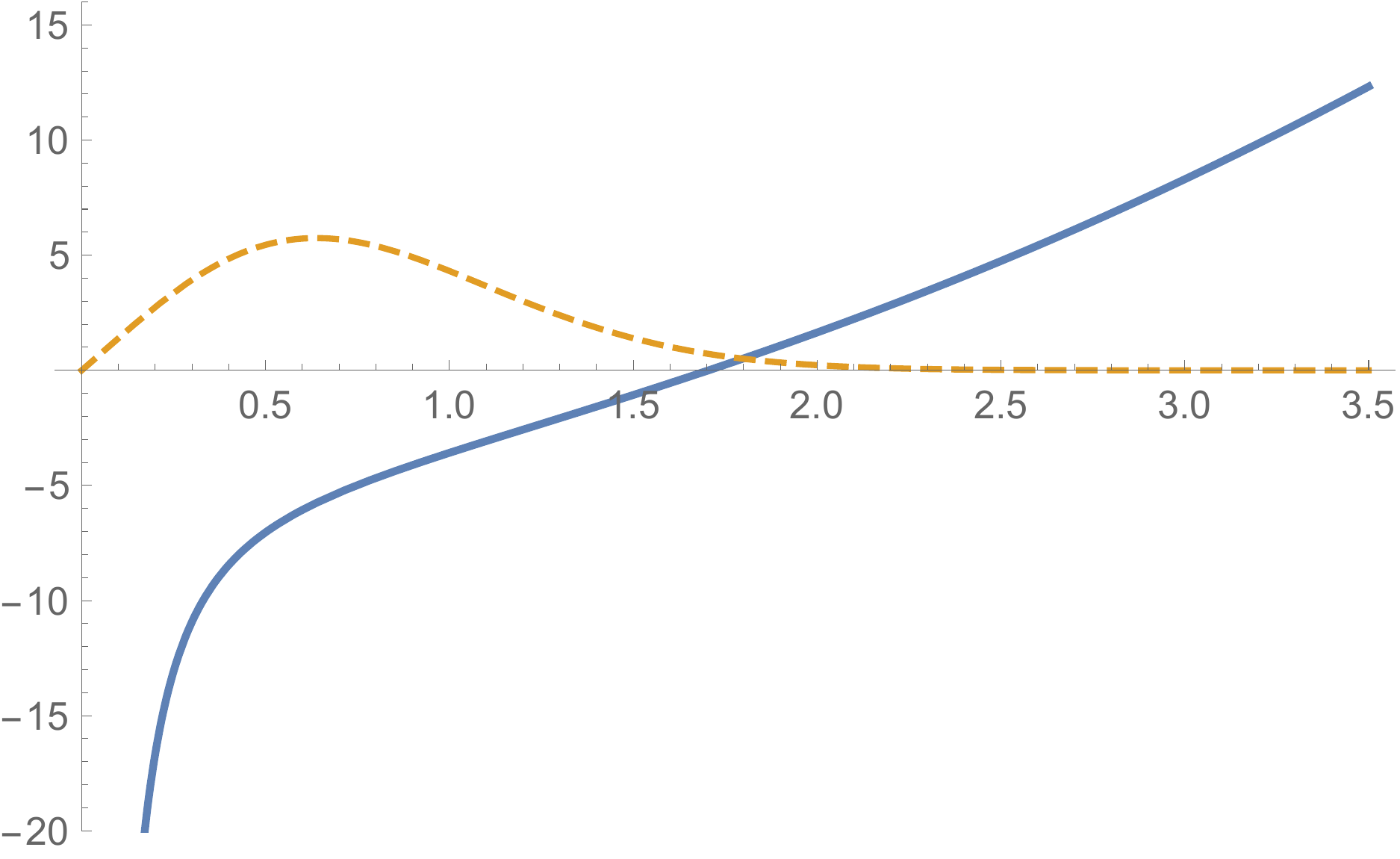}
\caption{Plots for the case  $c=-\sqrt{2}$ of the potential $V_1(x)$ (solid line) and probability density function $|\chi_p(x)|^2$ (dashed line) for $E_1=(3+\sqrt{5})/2$ (left) and $E_1=(3-\sqrt{5})/2$ (right).}
\label{fig12b}
\end{figure}

If we choose $b=1$, that is, $\rho=x$, then we may say that for the two values of the parameter $E$ shown in \eqref{e12} ($E_1, E_2$), the corresponding potential $V_1(x)$ in \eqref{potb1} has zero energy eigenvalue, as mentioned before.
A plot of the potentials and the probability density functions $|\chi_p(x)|^2$ can be seen in Figure~\ref{fig12}. The plot of corresponding potentials obtained with $c=-\sqrt{2}$ is shown in Figure~\ref{fig12b}.

\begin{figure}[ht]
\centering
\includegraphics[width=0.47\textwidth]{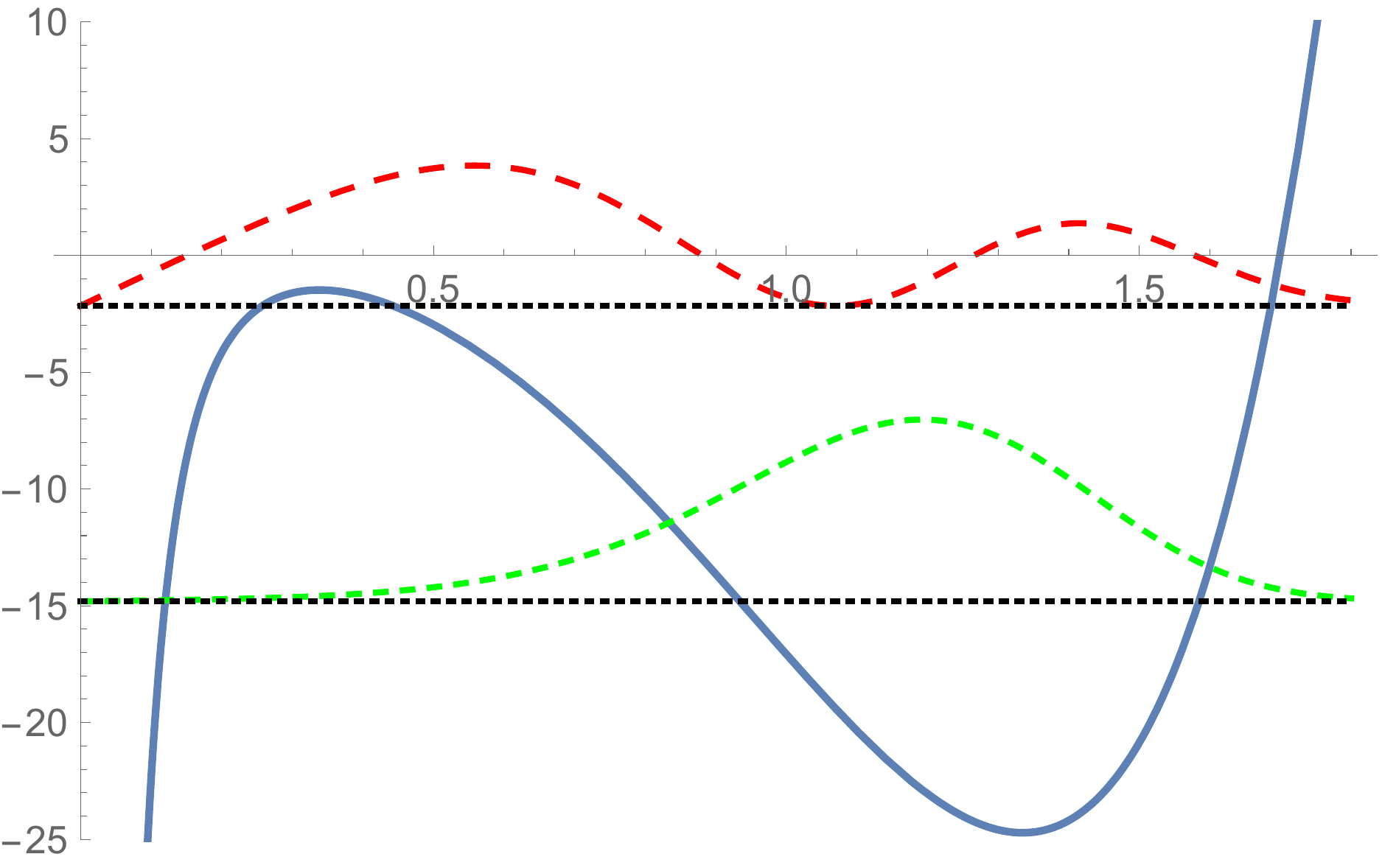}
\caption{
Plots for $c=\sqrt{2}$ of the potential $\tilde{V}_{1/2}(x)$ (solid line) and two probability density functions $|\chi_p(x)|^2$ (dashed lines) placed at the eigenenergy levels $\varepsilon_\pm=-2\sqrt{2} (3\pm \sqrt{5})$ (dotted lines).}
\label{fig3}
\end{figure}

\begin{figure}[ht]
\centering
\includegraphics[width=0.47\textwidth]{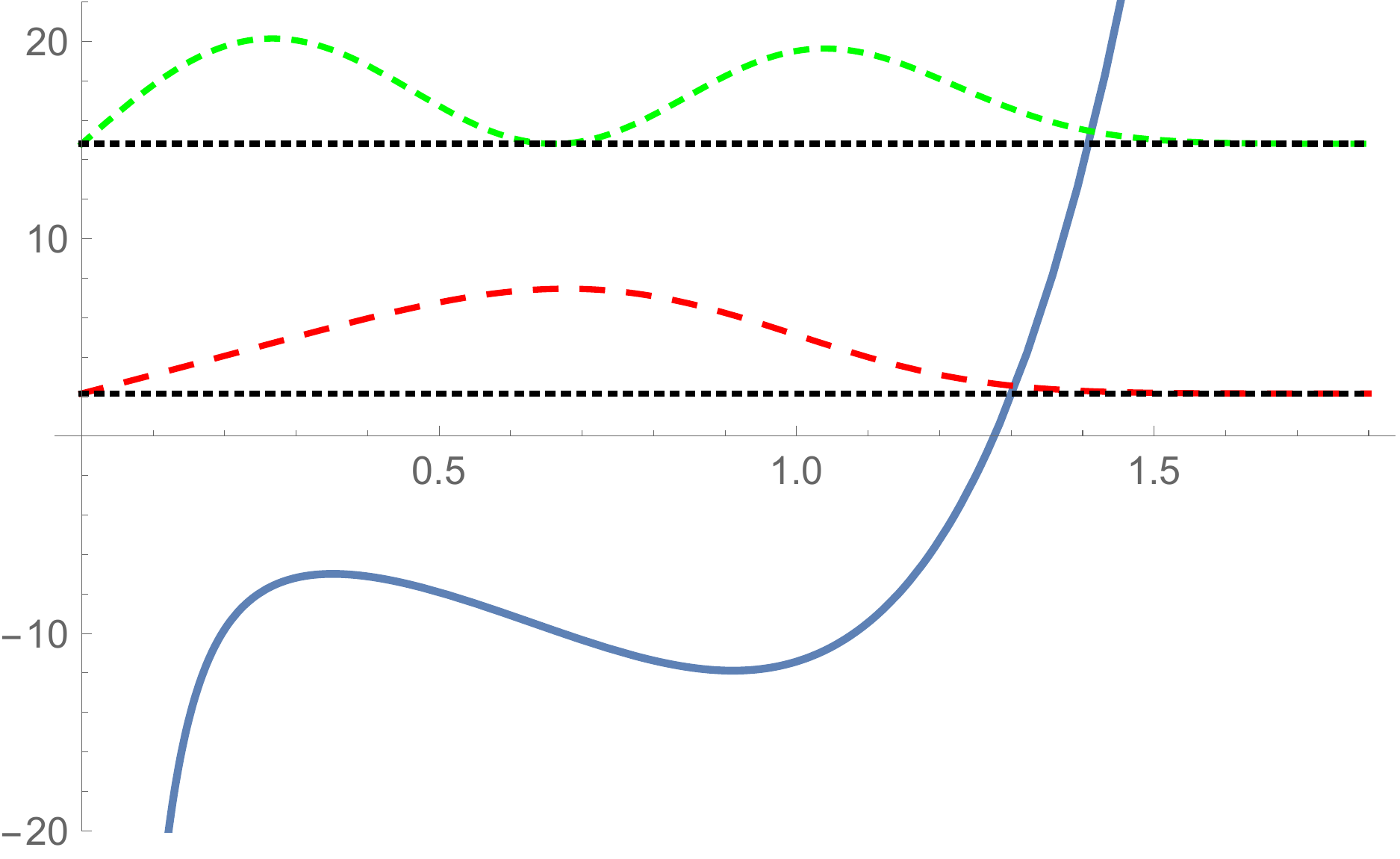}
\caption{
Plots for $c=-\sqrt{2}$ of the  potential $\tilde{V}_{1/2}(x)$ (solid line) and two probability density functions $|\chi_p(x)|^2$ (dashed lines) placed at the eigenenergy levels $\varepsilon_\pm=2\sqrt{2} (3\pm \sqrt{5})$ (dotted lines).}
\label{fig3b}
\end{figure}

If we choose $b=1/2$, that is $x=\sqrt{\rho}$, then the two values of $E$ in \eqref{e12} provide  energies of the potential $\tilde{V}_{1/2}(x)$, which turn out to be $\varepsilon_{\pm}=-2\sqrt{2} (3\pm\sqrt{5})$. A plot of the unique potential $\tilde{V}_{1/2}(x)$ and the probability density functions $|\chi_p(x)|^2$ can be seen in Figure~\ref{fig3} for $c=\sqrt{2}$
and Figure~\ref{fig3b} for $c=-\sqrt{2}$.

\subsection{Case $m=2$, $\ell=3$}

The three dimensional subspace in this case is $W_{3,2}$ and the restricted Hamiltonian $H_{3,2}$, as it was shown in Section~\ref{examples}.
%
%is generated by the vectors
%$$
%\{ |2,1,0 \rangle, |0,3,2 \rangle, |1,2,1\rangle\}.
%$$
%The matrix representation of the Hamiltonian in this subspace is ($\omega_1=\omega_2=\omega_3=1$):
%$$
%H=\left(
%\begin{array}{ccc}
%3&0&2\\
%0&5&\sqrt{6}\\
%2&\sqrt{6}& 4
%\end{array}
%\right),
%$$
The eigenvalues and eigenfunctions in this 3D subspace are:
\be
|\psi_p \rangle= \gamma_{p,1} |2,1,0 \rangle+ \gamma_{p,2} |0,3,2 \rangle +\gamma_{p,3}  |1,2,1\rangle ,\quad p=1,2,3,
\ee
with
\be\label{e123}
\begin{tabular}{|c|c|r|r|r|}
\hline 
&&&& \\ [-1.9ex]
  $p$ &     $E_p$         & $\gamma_{p,1}$\phantom{\quad} & $\gamma_{p,2}$\phantom{\quad} &   $\gamma_{p,3}$\phantom{\quad} \\ [0.9ex]
\hline 
&&&& \\ [-1.9ex]
1&$7.40405$  &      0.30313      &      0.68015      &   0.66750\\ [0.9ex]
\hline 
&&&& \\ [-1.9ex]
2&$ 3.81763$  &      0.72847      &     -0.61627      &   0.29781\\ [0.9ex]
\hline 
&&&& \\ [-1.9ex]
3&$ 0.77833$  &     -0.61437      &     -0.39598      &   0.68246\\ [0.9ex]
\hline
\end{tabular}
\ee
We can write those eigenfunctions in terms of the variables $\rho, s, t$ as follows:
\be
\psi_p \equiv s^3 t^2\ \rho^3 \left( 2 \gamma_{p,1} + \frac4{\sqrt{6}} \gamma_{p,2}\ \rho^2 + \frac4{\sqrt{2}} \gamma_{p,3}\ \rho \right).
\ee
The wavefunction is
\be
\chi_p(x)= e^{W(x)}\ \left( 2 \gamma_{p,1} + \frac4{\sqrt{6}} \gamma_{p,2}\ \rho(x)^2 + \frac4{\sqrt{2}} \gamma_{p,3}\ \rho(x) \right).
\ee

\begin{figure}[ht]
\centering
\includegraphics[width=0.32\textwidth]{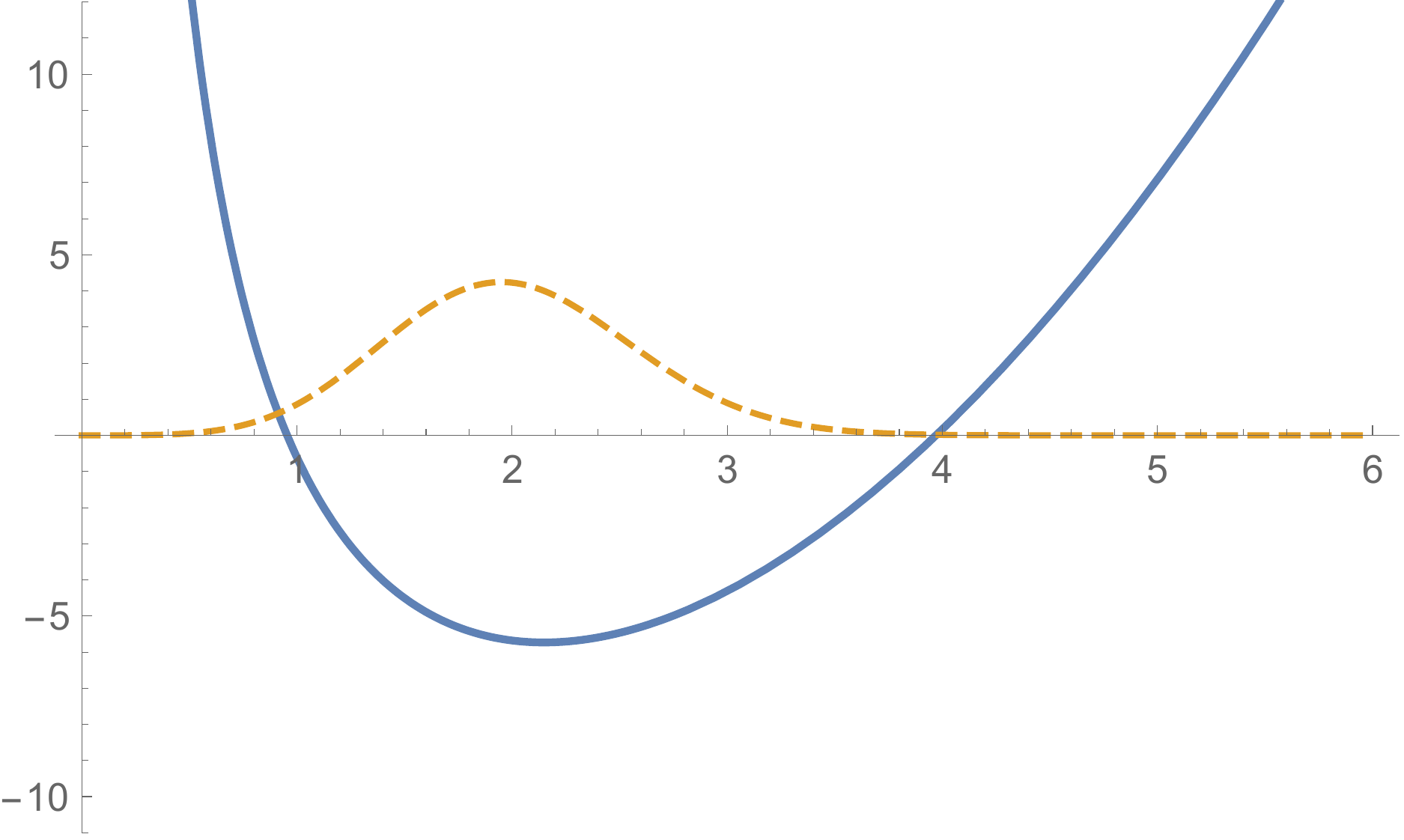}
\includegraphics[width=0.32\textwidth]{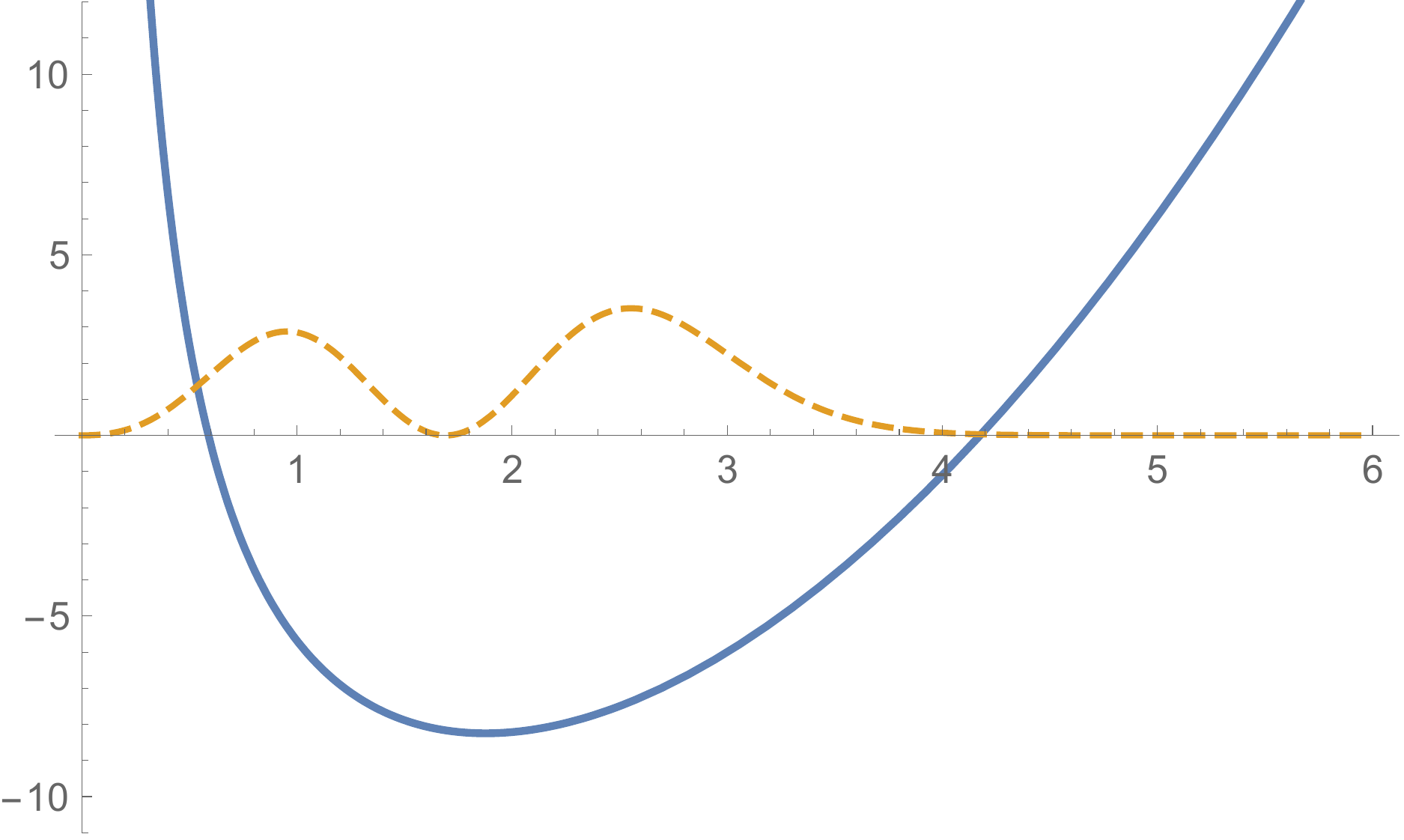}
\includegraphics[width=0.32\textwidth]{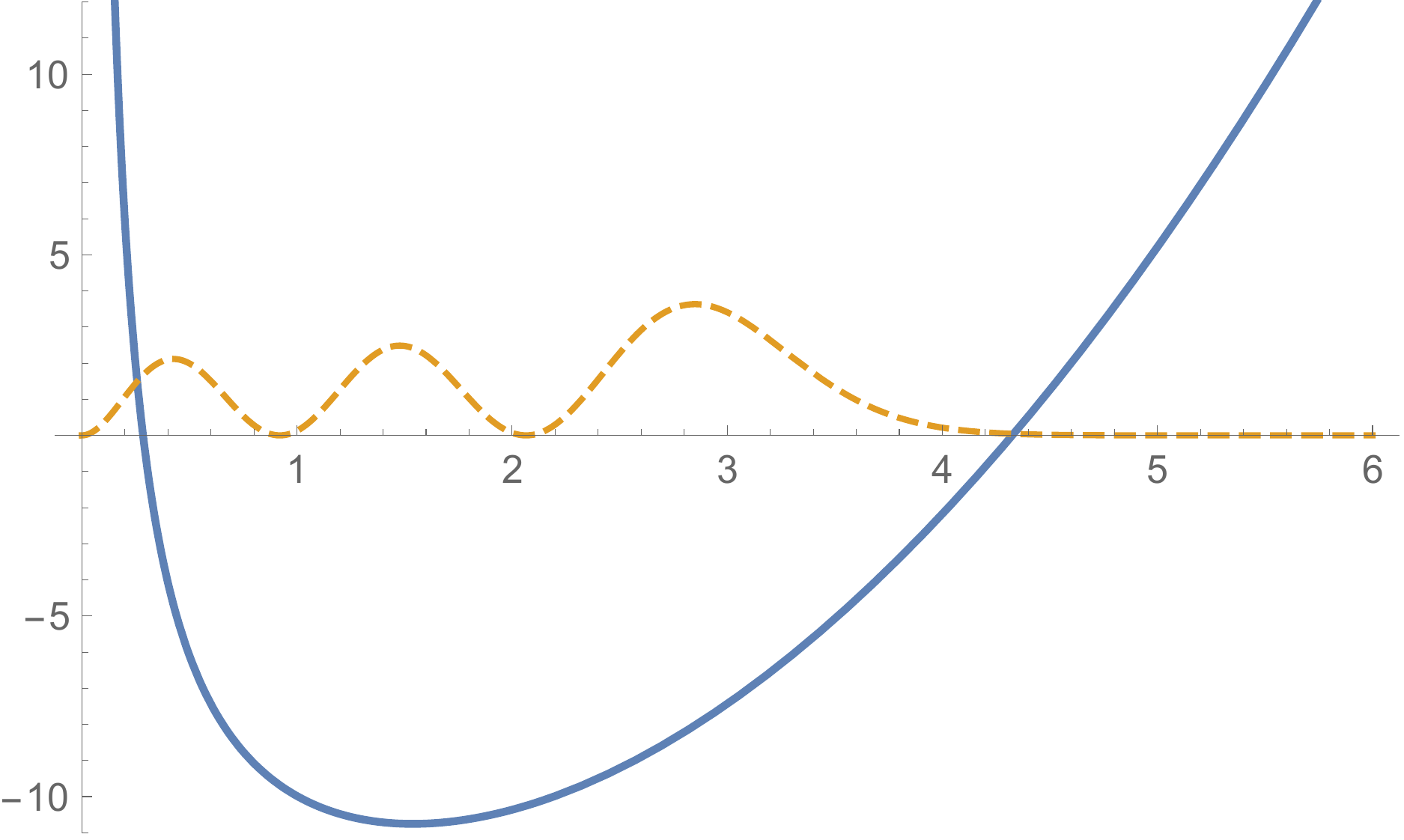}
\caption{The potentials   $V_1(x)$ (solid line) and probability density functions $|\chi_p(x)|^2$ (dashed lines) for $E_1= 7.40405$  (left), $E_2= 3.81763$ (center),  and $E_3=0.77833$  (right).}
\label{fig456}
\end{figure}

If we choose $b=1$, that is, $\rho=x$, then we say that the three values of the parameter $E$  shown in \eqref{e123}, ($E_1, E_2, E_3$), are such that the corresponding   potentials $V_1(x)$ have zero energy eigenvalue.
A plot of the  potentials and the probability density functions $|\chi_p(x)|^2$ can be seen in Figure~\ref{fig456}.

\begin{figure}[ht]
\centering
\includegraphics[width=0.5\textwidth]{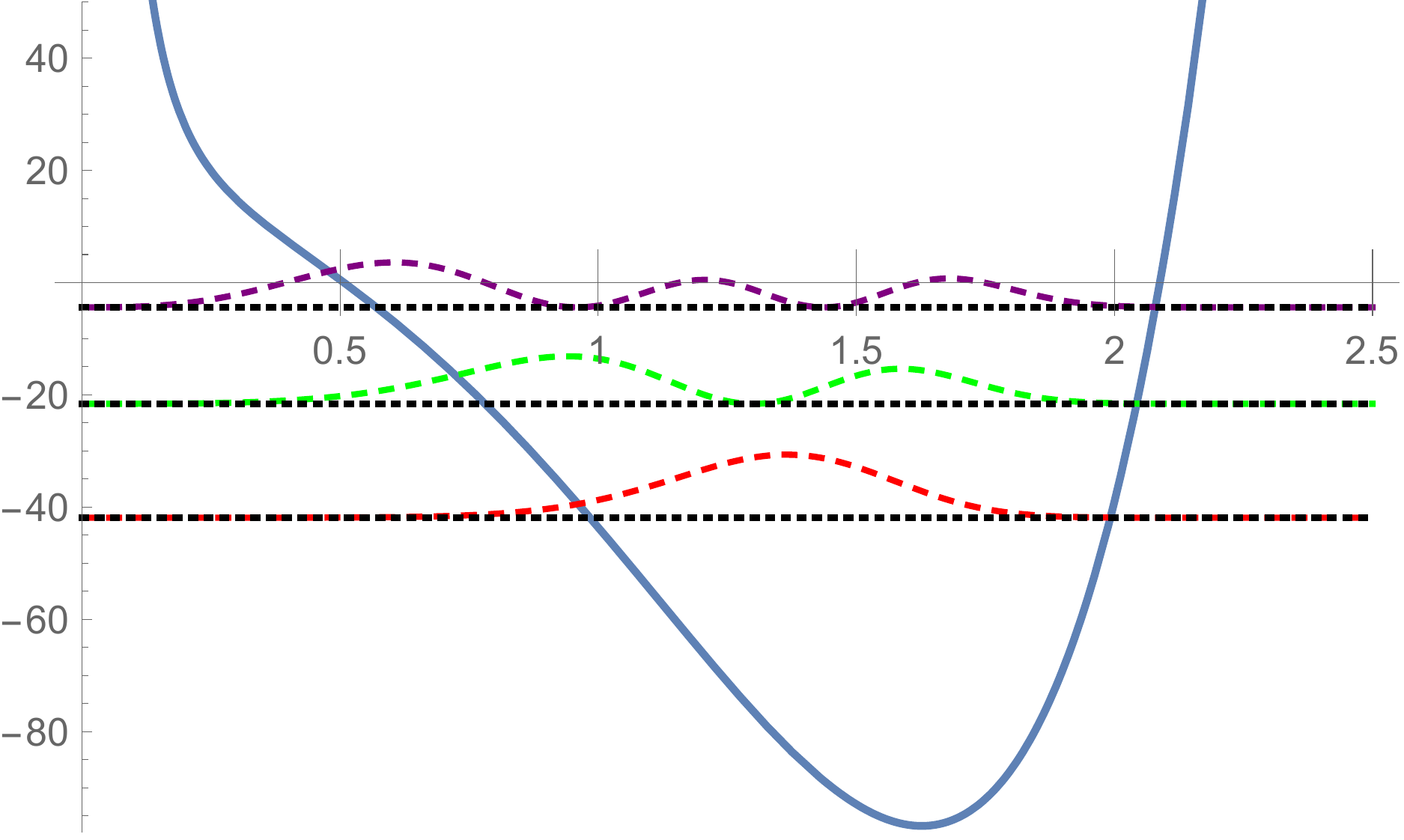}
\caption{The potential   $\tilde{V}_{1/2}(x)$ (solid line) 
and probability density functions $|\chi_p(x)|^2$ (dashed lines) placed 
at the energy levels   $\varepsilon_p=-4\sqrt{2} E_p$ (dotted lines).}
\label{fig7}
\end{figure}

On the other hand, if we choose $b=1/2$, that is $x=\sqrt{\rho}$, then the three values of $E$ in \eqref{e123} provide the eigenvalues of the Schr\"odinger equation  associated to $\tilde{V}_{1/2}$, which turn out to be $\varepsilon_p=-4\sqrt{2} E_p$.
A plot of the potential and the probability density functions $|\chi_p(x)|^2$ can be seen in Figure~\ref{fig7}.

\section{Conclusions}

In this paper we have presented a method to connect quantum optic models to 
quasi-exactly solvable Schr\"odinger equations in  a straightforward way. This method consists of the following steps:
\begin{itemize}
\item[1.] 
Characterise the symmetries of the quantum optical model and the invariant subspaces determined by their eigenvalues.

\item[2.] 
Set the quantum optical Hamiltonian in the standard differential realisation {\it a la} Fock-Bargmann. Then, separate the differential equation by means of a new set of variables by using the symmetries, to get an ordinary differential equation in one variable.

\item[3.] 
This differential equation can be rewritten as a Schr\"odinger equation which will be quasi-exactly solvable.
\end{itemize}
In this process, we start with just one Hamiltonian specified by the scaled frequencies $\omega_1, \omega_2, \omega_3$,  but its restriction to invariant subspaces ${\cal W}_{\ell, m}$ leads to a family of quasi-exactly solvable differential equations labeled by $\ell, m$.
This method is quite flexible: 

(i) For each value $\kappa={\rm min}(\ell,m)$ there are an infinite number of subspaces 
${\cal W}_{\ell, m}$ with the same dimension and therefore,
of quasi-exactly solvable potentials. 

(ii) Once fixed $\ell,m$, the last change of variable $x=\rho^b$, allows to find different types of potentials. 

(iii) This method can be applied to any number of interacting modes or atoms with different number of levels (work along this line is in progress). 

In this paper we have worked out two simple examples, the first one is a two dimensional subspace ${\cal W}_{1,1}$. For the value $b=1$ we have two potentials and we get one eigenfunction for each of these potentials. These eigenfunctions
correspond to the ground state and to the first excited state
(both having zero energy) of the respective potential, as shown in
Figs.~\ref{fig12} and \ref{fig12b}. When $b=1/2$ we  basically obtain only one potential  
 $\tilde{V}_{1/2}$ with two different
eigenfunctions corresponding to the ground and the first excited states,
but their energies have negative values. In this example the centrifugal term is missing,
but we could have chosen other two-dimensional examples where the centrifugal term be present.
In the second example,  we dealt with the  three
dimensional subspace ${\cal W}_{3,2}$ and similar considerations apply: we get either a list of three potentialseach one with a zero energy wavefunction, or one potential with three different
solutions for the first energy levels.

In conclusion, we have been able to connect a physical model in quantum optics,
the trilinear Hamiltonian, with a large list of Schr\"odinger type  systems 
which are quasi-exactly solvable, some of them with a clear physical meaning including
the quarkonium potential, or the sextic harmonic oscillator.

\section*{Acknowledgements}

This work was partially supported by the Spanish MINECO (MTM2014-57129-C2-1-P), Junta de Castilla y Le\'on and FEDER projects (BU229P18, VA057U16, and VA137G18). 
T. Mohamadian gratefully acknowledges financial support from the Ministry of Science, Research and Technology of Iran. She also would like  to thank the members of the Mathematical Physics Research Group of the University of Valladolid for their kind assistance and hospitality.

%\bibliographystyle{IEEEtranN}
%\bibliography{References}

\end{document}